\documentstyle[preprint,aps]{revtex}
\begin{document}
\draft
\tightenlines
\title{{Spectrum of elementary and 
collective excitations in the dimerized S=1/2 Heisenberg chain with frustration}}
\author{P.V. Shevchenko$^{a}$, V.N. Kotov$^{b}$, and O.P. Sushkov$^{c}$}
\address
{School of Physics, The University of New South Wales,
Sydney 2052, Australia}
\date{11 December 1998}
\maketitle
\begin{abstract}
We have studied the low-energy excitation spectrum of  a dimerized and frustrated 
antiferromagnetic Heisenberg chain. We use
an  analytic approach,
  based on a description of the excitations as triplets above a strong-coupling
 singlet ground state.
The quasiparticle spectrum is calculated by treating the excitations as a dilute Bose gas with infinite on-site repulsion. 
Additional singlet (S=0) and triplet (S=1) modes are found as two-particle bound states of the elementary triplets. We have also calculated  the 
contributions of the elementary and collective excitations into the  spin structure
 factor. Our
 results are in excellent agreement with exact diagonalizations and dimer 
series expansions data as long as the  dimerization parameter 
$\delta$ is not too small ($\delta>0.1$), i.e. while the 
elementary triplets can be treated as localized objects.
\end{abstract}


\baselineskip 0.6cm
\section{Introduction}
The  properties of a variety of recently discovered quasi one-dimensional materials can 
be described by the Alternating Heisenberg Chain (AHC) model. 
The Hamiltonian of the model reads:
\begin{equation}
\label{Ham}
H=J\sum_i [(1+\delta(-1)^i){\bf S}_i .{\bf S}_{i+1}+\alpha{\bf S}_i .{\bf S}_{i+2}]
,\end{equation}
where $i$ denotes the sites of a chain with length $N$ and ${\bf S}_i$ are $S=1/2$
 spin operators. The parameter $\alpha$ is the next-nearest neighbor coupling,
leading to frustration,  and $\delta$ is the dimerization. We assume $J>0$.

The AHC is  realized in nature in many materials that have two important but structurally inequivalent superexchange paths that are spatially linked, so that an alternating  spin-spin interaction results. Representative examples of  materials of this type  are $(VO)_2P_2O_7$ \cite{Garret}  and various aromatic free-radical compounds \cite{Nordio}. Alternating chains may also arise as a result of  spontaneous dimerization, due to dynamical spin-phonon coupling. 
Examples include the recently discovered inorganic spin-Peierls compounds  
  $CuGeO_3$ \cite{Hase} and $\alpha'-NaV_2O_5$ \cite{Isobe}.

The AHC model (\ref{Ham}) is a straightforward generalization of the uniform Heisenberg chain, which is the most widely studied quantum spin system. The uniform $S=1/2$ chain has a gapless excitation spectrum with a known dispersion relation and a rather complicated ground state which is characterized by strong quantum fluctuations, making it highly unstable to perturbations. The AHC generalizes the uniform chain by alternating the spin-spin interaction between two values, $J(1+\delta)$ and $J(1-\delta)$, and including next-nearest neighbor coupling (frustration) $\alpha J$. Since the Hamiltonian (\ref{Ham}) is rotationally invariant with respect to spin, the total spin is a good quantum number, and the ground state is a spin singlet. The translational symmetry however is broken by the dimerization, and the resulting system has a gap to the first excited state with $S=1$.
 The model has a rather complicated spectrum of states at higher energies, including multimagnon continua and bound states.

Recent neutron scattering studies of $(VO)_2P_2O_7$ carried out by Garrett {\it et al.}
\cite{Garret,Garret1,Garret2}  show that this material can be described well by the
alternating spin chain model. In addition, these measurements 
provide evidence for the existence of a
 two-magnon triplet ($S=1$) bound state. Similar studies of the spin-Peierls
 material  $CuGeO_3$ by   Ain {\it et al.}\cite{Ain}, in combination with 
 Raman spectroscopy \cite{Paul}, suggest that a singlet $S=0$ bound
 state is present in this material. 
These experiments have strongly motivated the  theoretical studies of the
low-energy excitations  of the AHC model, since the existence of magnetic bound states 
 is one of its characteristic features. 
More generally,  bound states seem to appear in all
low-dimensional dimerized quantum antiferromagnets which have a 
  gapped triplet spectrum above a singlet ground state \cite{Sushkov1}.

Besides the relevance to real compounds, the model (\ref{Ham}) is interesting from  theoretical point of view since it contains two independent mechanisms for spin gap formation. Subject of a special interest is the nature of excitations. At $\delta=0$ the model  
exhibits two phases, separated by a critical value of frustration $\alpha_c=0.2411$, known from numerous studies \cite{Okamoto}. For $\alpha<\alpha_c$ the ground state
is similar to that of the uniform Heisenberg chain and frustration is irrelevant in
 this parameter regime.
 The elementary excitations are massless 
unconfined  spinons with spin S=1/2 \cite{ft}. At $\alpha=\alpha_c$ there is a transition into a
phase with 
 spontaneously dimerized two-fold degenerate ground state. The spectrum acquires a gap and the elementary excitations are massive spinons \cite{Haldane,Shastry,Chitra}. On the other hand Haldane \cite{Haldane} has shown  that for any $\delta\neq 0$  the spinons become confined into
 triplet $S=1$ excitations with a gap in the spectrum.  
Interactions between the triplets can lead to additional massive singlet, $S=0$, and  triplet, $S=1$, excitations below the continuum \cite{Uhrig,Shastry} and even a sequence of further massive excitations \cite{Affleck1}. As $\delta\rightarrow 0$  the size of the elementary triplet
(or, equivalently, the spinon radius of confinement),  $r_0$, is known to scale as \cite{Uhrig1,Affleck2}: $r_0\sim \delta^{-2/3}$ for $\alpha<\alpha_c$ and  $r_0\sim \delta^{-1/3}$ for $\alpha>\alpha_c$. Therefore
 the triplets become nonlocal objects  for small dimerization, ultimately giving way to
 completely unconfined spinons in the strict limit $\delta=0$.

The excitation spectrum of the AHC has been extensively studied by 
a variety of numerical methods.
 Most  recent work includes:  Exact Diagonalizations (ED) carried out
 by Bouzerar, Kampf, and Japaridze \cite{Bouzezar} and Barnes, Riera and Tennant \cite{Barnes},
dimer series expansions by Singh and Weihong \cite{Singh}, and density matrix renormalization 
group (DMRG) studies by Sorensen {\it et al.} \cite{Affleck}.
However there are only few analytical works studying  the AHC model.
 Uhrig and Schulz\cite{Uhrig} have used the continuum limit field theory approach
 and predicted the possibility of singlet (S=0) and triplet (S=1) two-magnon bound states below
the two-particle continuum in the dimerized phase ($\delta>0$). Affleck  \cite{Affleck1} and Uhrig {\it et al.}\cite{Uhrig1} developed a description, based on   the soliton picture in which the elementary excitations are treated as
confined $S=1/2$ solitons (spinons). It was demonstrated that $S=1,0$ soliton anti-soliton bound states are formed  and their number increases
 as $\delta\rightarrow 0$.  
 Bouzerar and Sil \cite{Bouzezar1} have studied the excitation spectrum of the dimerized phase using the Bond Operator Technique (BOT) introduced by Chubukov\cite{Chubukov} and Sachdev and Bhatt\cite{Sachdev}, in combination with the   Brueckner approach developed by
 Kotov, Sushkov, Weihong, and Oitmaa \cite{Sushkov}. The results of Ref.\cite{Bouzezar1} are mainly inconsistent with the ED data. They are also inconsistent with our results in spite of the fact that we use the same approach. The reason for this discrepancy, as will become clear from
 our presentation below, is that the treatment of  Ref.\cite{Bouzezar1}   
is quite incomplete and misses several important 
technical aspects of the problem. 

The aim of the present work is to investigate theoretically 
the excited states of the dimerized phase ($\delta>0$) of the AHC by using the 
analytic Brueckner approach. Previous applications of the  Brueckner approach include the two-layer
 Heisenberg model \cite{Sushkov,Shevchenko}, the quantum spin-ladder \cite{Sushkov1} as well as the spin ladder with frustration \cite {Kotov}.  In all
cases excellent agreement was achieved between the theoretical
 and numerical spectra.
The Brueckner approach is based on the description of the excitations as triplets above a strong coupling singlet ground state.
 In essence it represents an effective way of taking into account
 the hard-core constraint, which has to be imposed on the triplets,
 by treating the excitations as  a dilute Bose gas with 
infinite on-site repulsion.  
 The method is  valid while  the  on-site density of excitations (triplets)
$n_{i}$ is small
 and under the assumption that the elementary triplets are well localized
 objects. In spite of the fact that the
  on-site density in the AHC is small even 
close to the point 
 $\delta=0$ ($n_i\approx 0.08$ for $\delta=0.01,\alpha=0$), we find that
 our approach fails at $\delta\sim 0.1$ because the triplets become nonlocal objects. In the region  $\delta > 0.1$ the present method provides an excellent
 quantitative description of the excitation spectrum. 

The rest of the paper is organized as follows.
 In Sec.II 
 we find the system of self-consistent equations describing the AHC with frustration and  calculate the one-particle (elementary triplet) spectrum for
a wide range of parameters. 
 Sec.III describes the 
 additional singlet and triplet modes which appear below the
 two-magnon continuum. These excitations are bound states of two
 elementary triplets. In Sec.IV  we calculate 
 the contributions of the  elementary and the additional triplet into
 the neutron scattering structure factor. Sec.V contains our conclusions. 

\section{Formulation of the problem and quasiparticle spectrum.}

In order to analyze the excitation spectrum of (\ref{Ham}) it is
 convenient to adopt the strong-coupling viewpoint. In the
limit $\delta=1, \alpha=0$ the  ground state consists of non-overlapping spin singlets $|GS>=|1,0>|2,0>|3,0>...$, where $|i,0>=\frac{1}{\sqrt{2}}[
|\uparrow>_{2i}|\downarrow>_{2i+1}-|\downarrow>_{i}|\uparrow>_{2i+1}]$.
Here, and in all formulas below which involve triplet operators,
  the "generalized" site index $i$ represents the bonds with 
spin exchange $J(1+\delta)$. 
 Each singlet can be excited into a triplet state and therefore it is   natural to introduce a creation operator $t_{\alpha i}^{\dagger}$ for this excitation: $|i,\alpha>=t_{\alpha i}^{\dagger}|i,0>$, $\alpha=x,y,z$. The representation of the spin operators in terms of the triplets $t_{\alpha i}^{\dagger}$ was introduced in  \cite{Chubukov,Sachdev} and reads: 
\begin{equation}
\label{bond}
S_{\alpha,2i}=\frac{1}{2}(t_{\alpha i}+t^{\dagger}_{\alpha i}-i\epsilon_{\alpha \beta
\gamma}t^{\dagger}_{\beta i}t_{\gamma i}) 
, 
\end{equation}
$$
S_{\alpha,2i+1}=\frac{1}{2}(-t_{\alpha i}-t^{\dagger}_{\alpha i}-i\epsilon_{\alpha \beta
\gamma}t^{\dagger}_{\beta i}t_{\gamma i}) 
,$$
where $\epsilon_{\alpha \beta\gamma}$ is the fully antisymmetric tensor.
After application of this transformation to (\ref{Ham}), or, equivalently, after calculating the matrix elements of the ``hopping'' terms, we find
the effective Hamiltonian 
\begin{equation}
\label{ham1}
H=H_2+H_3+H_4+H_U,
\end{equation}
\begin{equation}
\label{h22}
H_2=J_\perp\sum^{N/2}_{i=1,\alpha}t^{\dagger}_{\alpha i}t_{\alpha i}+\frac{\lambda}{2}\sum_{\alpha i}^{N/2}(t^{\dagger}_{\alpha i}t_{\alpha i+1}+t_{\alpha i}^{\dagger}
t^{\dagger}_{\alpha i+1}+h.c.),
\end{equation}
\begin{equation}
\label{h33}
H_3=\frac{\nu}{4}\sum^{N/2}_{i=1,\alpha \beta \gamma} (i\epsilon_{\alpha \beta \gamma}[t^{\dagger}_{\alpha i}t^{\dagger}_{\beta i+1}t_{\gamma i+1}-t^{\dagger}_{\alpha i+1}t^{\dagger}_{\beta i} t_{\gamma i}]+h.c.),
\end{equation}
\begin{equation}
\label{h44}
H_4=\frac{\mu}{2}\sum^{N/2}_{i=1,\beta \gamma}[t^{\dagger}_{\beta i}t^{\dagger}_{\gamma i+1}t_{\gamma i}t_{\beta i+1}-t^{\dagger}_{\beta i}t^{\dagger}_{\beta i+1}t_{\gamma i}t_{\gamma i+1}],
\end{equation}
\begin{equation}
\label{hu}
H_U=\frac{U}{2}\sum^{N/2}_{i=1,\alpha\beta} t^{\dagger}_{\alpha i}t^{\dagger}_{\beta i}t_{\beta i}t_{\alpha i},\hspace{0.3cm} U\rightarrow \infty,
\end{equation}
where we have adopted the notation:
\begin{equation}
J_\perp=J(1+\delta), \lambda=-\frac{1}{2}J(1-\delta-2\alpha), \mu=\frac{1}{2}J(1-\delta+2\alpha), \nu=J(1-\delta).
\end{equation} 
The present notation for the coefficients is chosen following Ref.\cite{Kotov}.
 For the problem studied in  Ref.\cite{Kotov}
the Hamiltonian  represents a spin ladder  with frustration and the term $H_3$
  cancels out due  to the symmetry of the ladder with respect to interchange of
 its two legs. The coupling $J(1+\delta)$  corresponds to the
 rung exchange $J_{\perp}$ in the spin ladder model, and gives the
 on-site energy of the triplets (first term in Eq.(\ref{h22})).       
 The rest  of the terms in the Hamiltonian represent
 an effective hopping of the triplets,  
  spontaneous creation of two triplets  on nearby sites (second term in  
Eq.(\ref{h22})), as well as interactions between the triplets (Eq.(\ref{h33}) and Eq.(\ref{h44})).    
In addition, an  infinite on-site repulsion $H_U$ is introduced to take into account the hard-core constraint $t^+_{\alpha i}t^+_{\beta i}=0$, which
has to be satisfied on every site.
 This condition (only one triplet can be excited
 on a site) follows from quantization of spin and guarantees the
 one to one correspondence between the spin model Eq.(\ref{Ham}) and
 its effective triplet description Eq.(\ref{ham1}).

Let us first consider the Hamiltonian (\ref{ham1}) without the $H_3$ 
term. A detailed analysis of the spectrum was presented in
 Refs.\cite{Sushkov,Sushkov1,Kotov} an we will only summarize the results here.
The interaction $H_U$ gives the dominant contribution
to the renormalization of the triplet spectrum. It was demonstrated in \cite{Sushkov} that in the Brueckner approximation this renormalization is given by the normal self-energy operator
\begin{equation}
\label{en}
\Sigma^{Br}_n(k,\omega)=4\sum_q Z^{Br}_{q}v_q^2\Gamma^{Br}(k+q,\omega-\omega_q)
,\end{equation}
where the scattering amplitude is
\begin{equation}
\label{ver}
\Gamma^{Br}(q,\omega)=-\left(\sum_p\frac{Z^{Br}_{p}Z^{Br}_{q-p}u_p^2u_{q-p}^2}{\omega-\omega_p-\omega_{q-p}}\right)^{-1}.
\end{equation}
The  renormalized spin-wave spectrum $\omega_k$, the quasiparticle residue $Z_k$ and the Bogoliubov parameters $u_k$, $v_k$ can be found from the solution of the coupled Dyson equations for the normal $G_n(k,t)=-i<T(t_{k\alpha}(t)t^{\dagger}_{k\alpha}(0))>$ and anomalous $G_a(k,t)=-i<T(t^{\dagger}_{-k\alpha}(t)t^{\dagger}_{k\alpha}(0))>$ Green's functions (see Ref.\cite{Sushkov}):
\begin{equation}
\label{sp}
\omega_{ k}=Z^{Br}_{ k}\sqrt{\tilde A_{k}^2-\tilde B_{k}^2},
\end{equation}
\begin{eqnarray}
\label{ab}
&&\tilde A_{k}=J_{\perp}+\lambda\mbox{cos}{2k}+\Sigma^{Br}({ k},0)+
2\mu\mbox{cos}{2k}\sum_{q}Z^{Br}_{q}v_{q}^2\mbox{cos}{2q},\\
&&\tilde B_{k}=\lambda\mbox{cos}{2k}-2\mu\mbox{cos}{2k}
\sum_{q}Z^{Br}_{q} u_{q}v_{q}\mbox{cos}{2q},\nonumber
\end{eqnarray}
\begin{eqnarray}
\label{zuv}
&&Z^{Br}_{k}=\left(1-\left.\frac{\partial \Sigma^{Br}}{\partial \omega}
\right|_{\omega=0}\right)^{-1},\\
&&u^2_{ k},v^2_{k}=\left(\frac{Z^{Br}_{k}\tilde A_{k}}
{2\omega_{k}}\pm\frac{1}{2}\right).\nonumber
\end{eqnarray}
These equations also take  into account the 
quartic interaction, $H_4$ (\ref{h44}), 
in the lowest order, one-loop approximation.
For the sake of simplicity, in all of the above formulas, and everywhere from now on, we
omit the number of lattice sites. 
All  momenta take values inside the  Brillouin zone of the original
(non-dimerized) lattice $(-\pi/2<k\le \pi/2)$.
The discrete lattice sums can be converted into integrals by
 $\sum_{k} \rightarrow \pi^{-1}\int_{-\pi/2}^{\pi/2}dk$.

The basic approximation made in the derivation of $\Gamma^{Br}(q,\omega)$,
as discussed in Ref.\cite{Sushkov}, is the neglect of all anomalous scattering vertices, which are present in the theory due to the existence of anomalous Green's functions (i.e. quantum fluctuations).  
All anomalous contributions however are suppressed by the small parameter of the Brueckner approach: the density of triplet excitations $n_i
=\sum_{\alpha}<t_{i\alpha}^{\dagger} t_{i\alpha}>
=3\sum_q Z^{Br}_{q}v_q^2$. As follows from Eq.(\ref{en}) the
 Brueckner self-energy is first order in $n_{i}$, while
 it is easy to prove that all  anomalous contributions lead to
  higher powers of the density \cite{remark}.
 This observation justifies the Brueckner approximation (i.e. keeping
 only Eq.(\ref{en})), since  the triplet
density 
  is small throughout the disordered phase, even close to  $\delta=0$.
 We find that   
 $n_{i} \approx 0.0066$ for $\delta=0.6, \alpha=0$, increasing at decreasing dimerizetion
 $n_{i} \approx 0.066$ for $\delta=0.1,\alpha=0$.

Now consider the effect of the cubic interaction $H_3$, Eq.(\ref{h33}).
In order to incorporate the effect of $H_3$
self-consistently,  the normal and anomalous self-energies have to  be found
to a given order and subsequently  inserted into the system of coupled Dyson's equations for the anomalous and normal  Green's functions. The  poles of these Green's functions 
 determine the renormalized one-particle spectrum \cite{Kotov1}.
 This procedure however is rather involved and in the present work 
 we  will use only the lowest order perturbation theory result (i.e.
 we will keep one diagram only). 
 This is justified, since we find that the lowest order contribution 
is relatively small and therefore  $H_3$
 does not require a fully self-consistent
 treatment.
The cubic interaction Eq.(\ref{h33}), after Fourier:  $t_{\alpha j}=\sum_{k}t_{\alpha k}e^{2ikj}$,
and Bogoliubov: $t_{k\alpha}=u_ka_{k\alpha}+v_ka_{-k\alpha}^{\dagger}$ transformations can be rewritten as
\begin{equation}
\label{h3}
H_3=\frac{1}{2}\sum_{k_1,k_2,k_3=k_1+k_2}\epsilon_{\alpha \beta \gamma}\left(\Gamma_1(1,2,3)[a^{\dagger}_{k_1\alpha}a^{\dagger}_{k_2\beta}a_{k_3\gamma}+h.c]
\right. +
\end{equation}
\begin{equation}
+\left.\frac{1}{3}\Gamma_2(1,2,3)[a^{\dagger}_{-k_1\alpha}a^{\dagger}_{-k_2\beta}a^{\dagger}_{k_3\gamma}+h.c.]\right).
\end{equation}
We have defined 
$$
\Gamma_1(1,2,3)=(u_{k_3}D(1,2,3)-v_{k_3}C(1,2,3))\sqrt{Z^{Br}_{k1}}\sqrt{Z^{Br}_{k2}}\sqrt{Z^{Br}_{k3}}
,$$
$$
C(1,2,3)=\Gamma_0(1,2)v_{k_1}v_{k_2}+\Gamma_0(-3,2)u_{k_2}v_{k_1}+\Gamma_0(1,-3)u_{k_1}v_{k_2},$$
where $\Gamma_0(1,2)=-\nu\mbox{sin}(k_1-k_2)\mbox{cos}(k_1+k_2)$ and  $D(1,2,3)=C(1,2,3)$ with replacing all $u_{\bf k}$ by $v_{\bf k}$ and vice versa $\{u \leftrightarrow v\}$,
and $\Gamma_2(1,2,3)=-\Gamma_1(1,2,3)$ with replacing $\{u_{k_3}\leftrightarrow v_{k_3}\}
$. 
Only a normal Green's function $G_n=-i<T(a_{k\alpha}(t)a_{k\alpha}^{\dagger}(0))>$ exists for the physical operators $a_{k\alpha}$, and therefore the 
self-energy induced by $H_3$ is given, to lowest order,  by the two
 diagrams shown in Fig.1:
\begin{equation}
\label{se3}
\Sigma_3(k,\omega)=\sum_{q}\left(\frac{\Gamma_1^2(q,k-q,k)}{\omega-\omega_q-\omega_{k-q}}-\frac{\Gamma_2^2(q,k-q,k)}{\omega+\omega_q+\omega_{k-q}}
\right).\end{equation}
The  correction $\delta\omega_k$ to the one-particle spectrum is: $\delta\omega_k=\Sigma_3(k,\omega_k)$.
The self-energy (\ref{se3}) also gives a 
 correction to the quasiparticle residue:
$
Z^{(3)}_k\approx 1+\left.\frac{\partial \Sigma_3(k,\omega)}{\partial \omega}\right|_{\omega=\omega_k}
$. 

In order to find the spectrum, equations (\ref{en},\ref{ver},\ref{sp},\ref{ab},\ref{zuv})
have to be solved self-consistently for $\Sigma^{Br}(k,0)$ and $Z^{Br}_{k}$,
 which corresponds to an infinite re-summation of diagrams (solution of Dyson's equation), and leads to the renormalized spectrum $\omega_{k}$.
 Then the correction due to $H_3$ has to be added resulting in the
 spectrum $\Omega_k=\omega_k+\delta\omega_k$.
 In this approximation the quasiparticle residue is  $Z_k=Z^{Br}_kZ_{k}^{(3)}$.
In the limit $\nu,\lambda\ll J_\perp$  all expressions  can be easily 
evaluated
 analytically to leading order, and for example the corrections, due to   
$H_{3}$ are:
 $\delta\omega_k=-\frac{\nu^2}{2J_\perp}\mbox{cos}^2k$, and $Z_{k}^{(3)}=1-\frac{\nu^2}{2J_\perp^2}\mbox{cos}^2k$. Finally, combining the three-particle corrections with the results of Ref.\cite{Kotov} for the one-particle 
spectrum without $H_{3}$, we obtain in the limit $\lambda,\nu\ll J_\perp$:
\begin{equation}
\label{s0}
\frac{\Omega_k}{J_\perp}=1+\frac{\lambda}{J_\perp}\mbox{cos}2k+\frac{3}{4}\frac{\lambda^2}{J^2_\perp}-\frac{1}{4}\frac{\lambda^2}{J^2_\perp}\mbox{cos}4k-\frac{1}{2}\frac{\nu^2}{J^2_\perp}\mbox{cos}^2k
.\end{equation}
Higher orders also can be evaluated, similarly to  Ref.\cite{Kotov},
 but we shall not present the lengthy expressions here.

 Next we present  results obtained by numerical self-consistent solution of
 the Dyson's equations. 
 Plots of the triplet gap $\Delta=\Omega_{k=0}/J$ as a function of $\alpha$ at $\delta=0.4,0.2$ are presented in Fig.2. Our results agree quite well with the
 ED data \cite{Bouzezar}. 
 We also note that our results disagree with those presented in
Ref.\cite{Bouzezar1}. The main reason for this is that the 
 three-leg vertices $H_3$, Eq.(\ref{ham1}) were not taken into account in
Ref.\cite{Bouzezar1}. In figures 4a,b,c,d we also
 present spectra of the elementary triplet $\Omega_k$ for different parameter values. The agreement with the numerical data (when  available) is very good.

The approach presented here relies heavily on the effective
 description of the spin problem in terms of localized triplets.
 As mentioned in the Introduction, in the limit  $\delta\rightarrow 0$,
 for any $\alpha$, the spinons,  confined to form the triplets,
 become unconfined. Our calculation clearly breaks down at this point.
 We would like to emphasize however, that due to the power law
 dependence of the radius of confinement $r_{0}$ on $\delta$,
 namely $r_0\sim \delta^{-2/3}$ for $\alpha<\alpha_c=0.2411$ and
 $r_0\sim \delta^{-1/3}$ for $\alpha>\alpha_c$ \cite{Uhrig1,Singh},
 the triplets are well localized for $\delta > 0.1$. For
$\delta \sim 0.1$ the triplet size is about 2-3 lattice spacings.
 Therefore the strong-coupling approach in fact is expected
 to  describe 
quantitatively well a very wide range of parameter space. 

   Let us also mention that the  effective Hamiltonian Eq.(\ref{ham1})
 can be used even for $\delta < 0.1$, when $r_{0}$ increases
 strongly. However technically the problem becomes much more
 difficult  and truly non-perturbative in nature. 
 One way to deal with it is to develop a perturbation theory to high
order, e.g.  
 via the dimer series expansion \cite{Singh}. Physically 
 a finite $r_{0}$ leads to the creation of a "string" (with length
 $r_{0}$) of excited triplets. This string   
 leads to the formation of low-energy many-particle bound states
 which mix strongly with the one-particle excitations, causing
 ultimately (for $\delta \rightarrow 0$) 
the vanishing of the one-particle spectral weight.
  These effects can be taken into account 
by diagonalizing the Hamiltonian in Fock space (spanned by
 the one-magnon, two-magnon, etc. states).
A detailed  analysis however 
is beyond the scope of the present work \cite{Kotov2}.

\section{ Spectrum of collective excitations: two-magnon bound states.}

The quartic term Eq.(\ref{h44}) in the Hamiltonian Eq.(\ref{ham1}) leads to   attraction  between two elementary triplets. It has been already demonstrated in 
the case of a 
 spin ladder \cite{Sushkov1,Kotov} that this attraction is strong enough to form  additional singlet (S=0) and triplet (S=1) bound states below the two-particle continuum $E^{C}(Q)=\mbox{ min}_q(\Omega_{\frac{Q}{2}-q}+\Omega_{\frac{Q}{2}+q})$. Here we will calculate the energies of 
these additional singlet and triplet states
 in the AHC model.  Such bound states have already  been 
 observed  in the ED calculations \cite{Bouzezar,Barnes} of the AHC.

Consider the scattering of the two triplets: $q_1\alpha+q_2\beta\rightarrow q_3\gamma+q_4\delta$ and introduce the total, $Q$, and relative, $q$, momenta of the pair.
The  two-particle singlet and triplet wave functions are:
\begin{equation}
\label{singlet}
|\psi^{S}(Q)>=\frac{1}{\sqrt{6}}\sum_{q \alpha}\psi^S(Q,q)a^{\dagger}_{\alpha \frac{Q}{2}-q}a^{\dagger}_{\alpha \frac{Q}{2}+q}|0>
,\end{equation}
\begin{equation}
\label{triplet}
|\psi_{\alpha}^T(Q)>=\frac{1}{2}\sum_{q,\beta\gamma}\psi^T(Q,q)\epsilon_{\alpha \beta \gamma}a^{\dagger}_{
\frac{Q}{2}+q \beta}a^{\dagger}_{\frac{Q}{2}-q \gamma}|0>,
\end{equation}
where $\psi^{S,T}(Q,q)$ are determined from the Schr{\"{o}}dinger equation,
$H|\psi^{S,T}(Q)>=E^{S,T}(Q)|\psi^{S,T}(Q)>$ with  bound state energy 
$E^{S,T}(Q)$. From this equation one can readily derive the
 integral Bethe-Salpeter equation satisfied by the bound state
 wave-functions:
\begin{equation}
\label{bs}
[E^{S,T}(Q)-\Omega_{\frac{Q}{2}+q}-\Omega_{\frac{Q}{2}-q}]\psi^{S,T}(Q,q)=
\sum_p M^{S,T}(Q,q,p)\psi^{S,T}(Q,p)
.\end{equation}
Let us at first
 neglect the cubic interaction $H_3$ in the Hamiltonian (\ref{ham1}).
 Then 
the scattering amplitudes $M^{S,T}(Q,q,p)$ in the singlet and triplet channels
 are (see Ref.\cite{Kotov}):
$
M^S(Q,q,p)=(U-2\mu\mbox{cos}2q\mbox{cos}2p)$, 
 and $
M^{T}(Q,q,p)=-\mu\mbox{sin}2q\mbox{sin}2p
$, respectively. Eq.(\ref{bs}) should be solved with the substitution  
\begin{equation}
M^{S,T}(Q,q,p)\rightarrow \sqrt{Z^{Br}_{\frac{Q}{2}-p}}\sqrt{ Z^{Br}_{\frac{Q}{2}+p}}\sqrt{ Z^{Br}_{\frac{Q}{2}-q}}\sqrt{ Z^{Br}_{\frac{Q}{2}+q}}
u_{\frac{Q}{2}-q}u_{\frac{Q}{2}+q}u_{\frac{Q}{2}-p}u_{\frac{Q}{2}+p}M^{S,T}(Q,q,p)
\end{equation}
in order to take into account that the triplet excitation $a^{\dagger}_{k\alpha}$
 differs from
the  bare one $t^{\dagger}_{k\alpha}$, due to the Bogoliubov transformation and the quasiparticle residue.
We stress that the solution of the integral equation (\ref{bs}) in the singlet channel 
has to  satisfy the condition
 $\sum_{q}\psi^S(Q,q)=0$, due to the 
infinite on-site repulsion. A  Lagrange multiplier has to be introduced to enforce 
this condition.
The normalization constants of the  wave functions (\ref{singlet},\ref{triplet}) 
are chosen to satisfy the conditions: $\sum_{Q} |\psi(Q)|^2=
\sum_q|\psi(Q,q)|^2=1.$
The solution of Eq.(\ref{bs}) for the triplet and singlet bound states (\ref{singlet},\ref{triplet}) has already been found in \cite{Kotov} for  the 
Hamiltonian (\ref{ham1}) without the $H_3$ term. In the leading order
 in $\lambda/J_\perp,\mu/J_\perp$ the energies and the
 wave functions of the bound states are:
\begin{equation}
\label{es}
E_Q^S=2J_\perp-\mu(1+C_Q^2)+\frac{3}{2}\frac{\lambda^2}{J_\perp}-\frac{\lambda^2}{4J_\perp}\mbox{cos}2Q,
\end{equation}
$$\psi^S(Q,q)= \sqrt{2(1-C_Q^2)}\frac{\mbox{cos}2q+C_Q}{1+2C_Q\mbox{cos}2q+C_Q^2}
,
$$
\begin{equation}
\label{et}
E_Q^T=2J_\perp-\frac{\mu}{2}(1+4C_Q^2)+\frac{3\lambda^2}{2J_\perp}+\frac{\lambda^2}{4J_\perp}\mbox{cos}2Q,
\end{equation} 
$$\psi^T(Q,q)=\sqrt{1/2-2C_Q^2}\frac{\mbox{sin}2q}{\frac{1}{2}+2C_Q\mbox{cos}2q+2C_Q^2}
,$$
where we have defined $C_Q=\frac{\lambda}{\mu}\mbox{cos}Q$.

Next, we analyze how $H_{3}$ affects the two-particle spectra. 
 The contribution of the three-particle scattering into binding should change the form 
of Eq.(\ref{bs})
 since it leads to retardation and thus to a non-trivial frequency dependence in the Bethe-Salpeter equation.
In the present work we treat
$H_3$ as a perturbation and find  the second order corrections, $\delta E_Q^{S,T}$, to the two-particle bound state energies $E^{S,T}_Q$. 
The  correction in the singlet channel
 is given as a sum of the  diagrams, shown in Fig.3a,b,c, 
convoluted with the singlet wave function
\begin{equation}
\label{dsin}
\delta E_1^S(Q)=\sum_{p,q}\delta M_a^S(Q,q,p) \psi^S(Q,q)\psi^S(Q,p)
+
\delta E^S_b(Q)+\delta E^S_c(Q),
\end{equation}
with the vertex  
\begin{equation}
\delta M_a^S(Q,q,p)=-\frac{4\Gamma_1(\frac{Q}{2}-q,q-p,\frac{Q}{2}-p)\Gamma_1(\frac{Q}{2}+p,q-p,\frac{Q}{2}+q)}{E^S_Q-\omega_{q-p}-\omega_{\frac{Q}{2}+p}-\omega_{\frac{Q}{2}-q}}
\end{equation} and
\begin{eqnarray}
\label{e0}
&&\delta E^S_b(Q)=
2\sum_{p,q}\psi_S^2(Q,p)\frac{\Gamma_1^2(q,p-q+\frac{Q}{2},p+\frac{Q}{2})}{E^S_Q-\omega_{p-q+\frac{Q}{2}}-\omega_{\frac{Q}{2}-p}-\omega_q}
,\\
&&\delta E_c^S(Q)=2\sum_{p,q}\psi_S^2(Q,p)
\frac{\Gamma_2^2(q,p-q+\frac{Q}{2},p+\frac{Q}{2})}
{E^S_Q-\omega_{p-q+\frac{Q}{2}}-\omega_q-2\omega_{\frac{Q}{2}+p}-\omega_{\frac{
Q}{2}-p}}.
\end{eqnarray}
The correction to the triplet energy $E^T(Q)$ is given by the diagrams a,b,c,d,e in Fig.3,
  convoluted with the triplet wave function
\begin{equation}
\label{dtrip}\delta E^T_1(Q)=\sum_{p,q}(\delta M^T_a+\delta M^T_d+\delta M^T_e)\psi^T(Q,q)\psi^T(Q,p)
+\delta E_b^T(Q)+\delta E_c^T(Q)
\end{equation}
with the vertices:  $\delta M_{a}^T(Q,q,p)=\frac{1}{2}\delta M^S_a(Q,q,p)\{E^S(Q)\rightarrow E^T(Q)\}$, 
$$
\delta M_{d}^T(Q,q,p)=\frac{\Gamma_1(\frac{Q}{2}+p,\frac{Q}{2}-p,Q)\Gamma_1(\frac{Q}{2}+q,\frac{Q}{2}-q,Q)}{E^T_Q-\omega_Q}
,$$
$$
\delta M_{e}^T(Q,q,p)=\frac{\Gamma_2(\frac{Q}{2}+q,\frac{Q}{2}-q,Q)\Gamma_2(\frac{Q}{2}+p,\frac{Q}{2}-p,Q)}{E^T_Q-\omega_Q-\omega_{\frac{Q}{2}-q}-\omega_{\frac{Q}{2}+q}-\omega_{\frac{Q}{2}-p}-\omega_{\frac{Q}{2}+p}}
,
$$
and the terms:  $\delta E_{b,c}^T(Q)=\delta E_{b,c}^S(Q)\{\psi^S,E^S\rightarrow
 \psi^T,E^T\}$.

In the leading order in $\nu,\lambda\ll J_\perp$ the corrections (\ref{dsin},\ref{dtrip}) can be easily found analytically:
\begin{equation}
\label{mom}
\delta E_1^S(Q)=-\frac{\nu^2}{2J_\perp}\mbox{cos}^2Q
,\hspace{0.2cm}\delta E_1^T(Q)=-\frac{\nu^2}{2J_\perp}(1-\frac{1}{2}\mbox{cos}^2Q)
.\end{equation}
We observe that the contributions (\ref{dsin},\ref{dtrip})  
 have an appreciable effect on
 the two-particle energies  and  therefore  the cubic interaction (\ref{h33}) 
can not be neglected.

It is worth noting that Eqs.(\ref{bs},\ref{dsin},\ref{dtrip}) are not quite correct because they do not take fully into account the  hard-core constraint $t^{\dagger}
_{i\alpha}t^{\dagger}_{i\beta}=0$.
Let us consider the limit $\lambda,\nu\ll \mu\ll J_\perp$. A direct expansion in powers of
  $1/J_\perp$ applied to the Hamiltonian (\ref{ham1}) in coordinate space, gives the
following energies for the singlet and triplet bound states:

\begin{eqnarray}
\label{real1}
&&E^S_{real}=2J_\perp-\mu-\frac{\lambda^2}{2\mu}(1+\mbox{cos}2Q)+\frac{9\lambda^2}{8J_\perp}-\frac{\nu^2}{4J_\perp}(1+\mbox{cos}2Q),\\
\label{real2}
&&E^T_{real}=2J_\perp-\frac{\mu}{2}-\frac{\lambda^2}{\mu}(1+\mbox{cos}2Q)+\frac{9\lambda^2}{8J_\perp}+\frac{\nu^2}{8J_\perp}\mbox{cos}2Q
.
\end{eqnarray}
The hard-core constraint is completely taken into account in these formulas by
not allowing hopping processes leading to two triplets occupying the same site.
On the other hand the momentum space diagrammatic calculation 
 Eqs.(\ref{es},\ref{et}) and the three-particle correction
 (\ref{mom}) result in
\begin{eqnarray}
\label{mom1}
&&E^S_{mom}=2J_\perp-\mu-\frac{\lambda^2}{2\mu}(1+\mbox{cos}2Q)+\frac{\lambda^2}{4J_\perp}(6-\mbox{cos}2Q)-\frac{\nu^2}{2J_\perp}\mbox{cos}^2Q,\\
\label{mom2}
&&E^T_{mom}=2J_\perp-\frac{\mu}{2}-\frac{\lambda^2}{\mu}(1+\mbox{cos}2Q)+\frac{\lambda^2}{4J_\perp}(6+\mbox{cos}2Q)-\frac{\nu^2}{2J_\perp}(1-\frac{1}{2}\mbox{cos}^2Q)
.
\end{eqnarray}
The physical reason for the difference between the 
  momentum space calculation and the  exact real space result  is in the
 additional blocking of virtual excitations which is not included in  Eqs.(\ref{bs},\ref{dsin},\ref{dtrip}). Let us consider one example of such additional blocking.  When we have
 a state with  one triplet, the quasiparticle blocks virtual excitations due to quantum fluctuations, $t^{\dagger}_{i\alpha}t^{\dagger}_{i+1\alpha}$, on two links. This is the physical origin of the third term in the one-particle dispersion (\ref{s0}).  When we have two 
excited
triplets and they are separated by more than one lattice 
spacing they block four links (this 
corresponds to the third terms in (\ref{es},\ref{et})).
However   when the two triplets are on nearest-neighbor sites they block only three links,
 which is energetically more favorable and leads to the increase of the binding energy.
This effect represents a non-potential contribution and is not taken into 
account by the integral equation (\ref{bs}).
The  additional contribution of the quantum fluctuations into binding  is analogous to the 
calculation of the Lamb shift in atomic physics.
It is clear from the above discussion that  binding due to blocking of
 quantum fluctuations  exists only 
 when the triplets are on nearest-neighbor sites.

We have also identified  several additional processes which  have to be described more
 accurately by inserting the  vertex  (\ref{ver}),  responsible for the infinite one-site
repulsion, into the
 diagrams presented in Fig.3. These diagrams should account correctly for
the hard-core repulsion
 between the particles in the initial and intermediate states.   
However it is very difficult to identify and calculate in momentum space
 all diagrams of this type. 
We will   follow a simpler, effective way to take into account these additional
 contributions. 
Comparing the coordinate space results (\ref{real1},\ref{real2}) with the momentum space calculations (\ref{mom1},\ref{mom2}) one can see that the latter 
 can be corrected by adding the following expressions to Eqs.(\ref{dsin},\ref{dtrip}): 
\begin{eqnarray}
\label{con}
&&\delta E^S_2(Q)=\frac{\lambda^2}{4J_\perp }\left[-\frac{3}{2}+\mbox{cos}2Q\right]\left|
\sum_{k}\sqrt{2}\mbox{cos}2k\psi^S(Q,k)\right|^2
,\\
\label{con1}
&&\delta E^T_2(Q)=\frac{3\lambda^2}{8J_\perp}\left[-1-\frac{2}
{3}\mbox{cos}2Q+\frac{\nu^2}{\lambda^2}\right]\left|\sum_{k}
\sqrt{2}\mbox{sin}2k\psi^T(Q,k)\right|^2
.\end{eqnarray}
In the above equations the expressions in the absolute value signs 
  give the probability amplitudes for two quasiparticles  to be on nearest-neighbor sites in the appropriate channel.
 Finally, to obtain the energies of the triplet and singlet bound states, the integral equation (\ref{bs}) should be solved first, and then the 
corrections $\delta E^{S,T}_1(Q)$ (\ref{dsin},\ref{dtrip}) and $\delta E^{S,T}_2(Q)$ (\ref{con},\ref{con1}) should be added.
Results obtained by a   self-consistent numerical 
 solution for the spectrum of the singlet and triplet bound states are 
presented in Fig.4 for different values  
 of the frustration, $\alpha$, and dimerization, $\delta$. 
We find  that there are no bound states in the vicinity of $q=0$ for zero frustration ($\alpha=0$). Singlet  and triplet bound states
 are always present in the vicinity of $q=\pi/2$, however the singlet typically
 exists within  a much wider momentum range.  
 At $q=0$ the singlet splits off
 from the continuum at any non-zero  $\alpha$, while the triplet 
 exists only above a certain value of frustration. 
 One can see also that, quite generally, the singlet is below the triplet, i.e.
 the binding in the singlet channel is stronger. 

We have also  calculated 
the binding energies, $\varepsilon^{S,T}(Q)=E^C(Q)-E^{S,T}(Q)$,   
 of the singlet and triplet bound states as a  function of  $\delta$ at
$Q=\pi/2$ for $\alpha=0$.
 The results are presented in Fig.5.
 There is excellent agreement  with the  ED data \cite{Bouzezar,Barnes} in
 the singlet channel even for $\delta=0.1$, 
whereas the agreement for the triplet  is not so good for $\delta<0.5$.
 We attribute this disagreement to the fact that 
the triplet binding energy is relatively small in comparison with  the singlet one for 
the whole range of  parameters ($\alpha,\delta$).
 Therefore taking into account the  three-particle scattering in simple perturbation theory is not as good approximation  for the  triplet as it is for the singlet.
 We have performed a more 
   accurate calculation  for the cases when the   binding energy 
 is very small. An 
improvement can be achieved by 
 adding the corrections $\delta M^{S,T}(Q,q,p)$ to the scattering amplitude $M^{S,T}(Q,q,p)$ into the Bethe-Salpeter integral equation (\ref{bs}) and further solving 
 the equation to find the binding  energies. 
Such "self-consistent" calculations are presented by the dashed lines in Fig.5 and one can see that the agreement  with  the ED data indeed becomes better for small energies.

There are several important points which have been overlooked in Ref.\cite{Bouzezar1},
ultimately   leading to incorrect results for the energies of the  bound states. The contribution of the constraint (\ref{hu}) has not been taken into account in the integral equation (\ref{bs}) 
resulting in non-zero binding energy of the singlet  at $q=0$ even for large $\delta$
and $\alpha=0$ (notice that our calculation always gives zero binding at this point). 
The contribution of the three-particle scattering (\ref{h33}) 
into binding has not been taken into account in Ref.\cite{Bouzezar1}
 and the effect given by the corrections (\ref{con},\ref{con1}) (Lamb shift) has not been included.
 In addition, the error in the calculation of the one-particle spectrum (see the previous section) 
  has propagated into the two-particle energies as well. 

Finally  we would like to compare  our results for the bound state energies
 with the lowest order perturbation theory results, presented in Ref.\cite{Barnes}.
The binding energy $\varepsilon^{S,T}(Q)=E^C(Q)-E^{S,T}(Q)$ at $Q=\pi/2,\alpha=0$ resulting from (\ref{real1},\ref{real2}) and (\ref{s0}) in the
 singlet channel, $\varepsilon^S(\frac{\pi}{2})/J=\frac{1}{2}(1-\delta)-\frac{17}{32}\frac{(1-\delta)^2}{1+\delta}$, is different from  
the result of  Ref.\cite{Barnes}: $\varepsilon^S(\frac{\pi}{2})/J=\frac{1}{2}(1-\delta)-\frac{14}{32}\frac{(1-\delta)^2}{1+\delta}$.
In the  triplet channel our result is
$\delta\varepsilon^T(\frac{\pi}{2})/J=\frac{1}{4}(1-\delta)-\frac{13}{32}\frac{(1-\delta)^2}{1+\delta}$,  which coincides with the formula presented in  Ref.\cite{Barnes}.
 
\section{Structure factor.}
The one-particle triplet and the triplet bound state
 can be observed in neutron scattering 
experiments. The inelastic neutron scattering cross-section is directly proportional to the dynamic structure factor:
\begin{equation}
S(q,\omega)=\frac{1}{2}\mbox{Im}\int e^{i\omega t}<\mbox{T}({\bf S}(q,t).{\bf S}(-q,0))>.
\end{equation}
Here we calculate the spectral weights of the elementary and the  additional triplets.
After the  bond operator (\ref{bond}) and Bogoliubov transformations, the on-site spin operator ${\bf S}_{i}$ can be written in momentum space as:
\begin{equation}
S_\alpha(-q,0)=-ie^{iq/2}\left[T_1(q)a_{q\alpha}^{\dagger}+\epsilon_{\alpha\beta\gamma}\sum_{k}T_2(k,q)a^{\dagger}_{k+q\beta}a^{\dagger}_{k\gamma}\right],
\end{equation}
where we have defined the
 vertices $T_1(q)=(u_q+v_q)\mbox{sin}\frac{q}{2}$ and
 $T_2(k,q)=u_{k+q/2}v_{k-q/2} \mbox{cos}\frac{q}{2}$.
Then, after averaging, the structure factor can be rewritten in the form  $S(q,\omega)=S_1(q)\delta(\omega-\Omega_q)+S_2(q)\delta(\omega-E^T_q)$ where 
\begin{equation}
\label{str1}
S_1(q)=\frac{1}{2}\left[\sqrt{Z^{(3)}}T_1(q)+2
\sum_{k}\frac{T_2(k,q)\Gamma_1(k+\frac{q}{2},-k+
\frac{q}{2},q)}{\omega_q-\omega_{k+\frac{q}{2}}-\omega_{k-\frac{q}{2}}}
\right]^2
\end{equation}
is the spectral weight of the elementary  triplet, formally represented  by the diagram in Fig.6a, and
\begin{equation}
\label{str2}
S_2(q)=\frac{1}{2}\left|2\sum_{k}T_2(k,q)
\psi^T(q,k)+
\right.\end{equation}
$$\left.
\frac{T_1(q)}{E^T_q-\omega_q-\Sigma_3(E^T_q,q)}
\sum_{k}\Gamma_1(k+\frac{q}{2},-k+\frac{q}{2},q)\psi^T(q,k)\right|^2
$$
is the contribution of the two-particle triplet bound state,  represented by the diagrams in Fig.6b. Note that in calculations of the spectral weights $S_1(q)$ and $S_2(q)$ the total Green's function, including the three particle scattering, should be used, i.e
 the corrected quasiparticle residue $Z=Z^{Br}Z^{(3)}$ and the total elementary spectrum $\Omega_k=\omega_k+\delta\omega_k$, should be substituted.
Along the disordered line $\lambda=0$  there are no quantum fluctuations
  (i.e. $v_k=0,u_k=1,Z^{Br}=1$), and
 the dependence of $S_1(q)$ and $S_2(q)$ upon $\delta$ is given by the three particle scattering contribution only. 
In the leading order in $\lambda\ll \mu,J_\perp$, we have calculated the spectral weights analytically:
\begin{eqnarray}
&&S_1(q)\approx \frac{1}{2}\mbox{sin}^2\frac{q}{2}(1-\frac{\lambda}{2J_\perp}\mbox{cos}2q-\frac{\nu^2}{4J^2_\perp}\mbox{cos}^2q-\frac{\lambda\nu}{J^2_\perp}\mbox{cos}q\hspace{0.1cm}\mbox{cos}^2\frac{q}{2})^2, \\ 
&& S_2(q)\approx \frac{1}{4}(\frac{\nu}{J_\perp}\mbox{cos}q\hspace{0.1cm}\mbox{sin}\frac{q}{2}+\frac{\lambda}{J_\perp-\mu/2}\mbox{cos}\frac{q}{2}\hspace{0.1cm}\mbox{sin}q)^2.
\end{eqnarray}
We have also performed a   self-consistent numerical evaluation of $S_1(q)$ and $S_2(q)$ using (\ref{str1},\ref{str2}), and the results are  presented in Fig.7a and Fig.7b.
For comparison we have also  plotted $S_1(q)$ 
 obtained by the dimer series expansion technique \cite{Singh}. 
The structure factor of the elementary triplet is in excellent agreement with the
dimer series  results even for $\delta=0.2$. From  Fig.7a, which shows the
 case $\lambda=0= 1-\delta-2\alpha$, 
it is clear that  for  small $\alpha$ (and large $\delta$),
 the structure factor of the bound state
triplet is much smaller than the one for the elementary triplet. 
However as $\delta$ decreases  the spectral weight of the bound state triplet
increases and eventually can even become equal to
  the one-particle contribution.

\section{Summary and conclusions.}

In summary, by using self-consistent diagrammatic analysis,
we have calculated the one-particle and the two-particle excitation
spectra of the dimerized and frustrated antiferromagnetic Heisenberg chain.
 The quasiparticle excitations are described as  a dilute, strongly-correlated
 Bose gas of triplets. The important parameter, which controls  the validity
 of the diagrammatic expansion and allows us to re-sum  effectively the
 most important diagrams is the density of triplets. The latter quantity
 is quite small for a wide range of parameters, making the technique, presented
 in this paper, effective for   dimerization $\delta>0.1$.
 For smaller dimerization the triplets become non-local objects and
  a more adequate picture would be one of loosely bound spinons. 
  We emphasize however that the region of applicability of the presented 
 technique is, perhaps even surprisingly, quite large. The reason for this
 can be traced to the fact that the quasiparticle weight decreases rather
slowly as $\delta$ decreases. Indeed, from    Fig.7a one can see that
 even for the quite small value   $\delta=0.1$ the residue is around $0.3$
(the  maximum is chosen to be 0.5). 
Even though this is true only in the special case without
 quantum fluctuations, it appears to be qualitatively correct even in the
 general case. Therefore the triplets are well localized objects with large
 quasiparticle residue even for rather small values of $\delta$.
We have found that our results for the spectrum are in excellent agreement with
the  numerical data, obtained by exact diagonalizations and dimer series
 expansions, for $\delta \approx 0.2$ and higher.

 We have presented a detailed diagrammatic analysis of the 
collective excitations and have found that singlet and triplet bound states of two
elementary  triplets
 generally exist below the two-particle continuum.
 In the absence of frustration there are no bound states in the
 vicinity of $q=0$, while both singlet and triplet ones exist at $q=\pi/2$.
 Finite frustration increases the binding energies of both excitations,
 making the singlet split off from the continuum at $q=0$ for any $\alpha\neq0$,
 while a finite  $\alpha$ is required for the triplet bound state to appear.
 We have also calculated the contribution of the elementary triplet and the triplet bound state into
 the structure factor, which is directly measurable in inelastic neutron 
 scattering experiments. We have found that for small $\alpha$ and large
 $\delta$ the bound state contribution is quite small, however it grows
 with decreasing  $\delta$ and can become equal to the
 quasiparticle contribution.

The technique used in this work is quite similar in spirit to
 the dimer series expansion, the major difference being that while
 we re-sum only the most important classes of diagrams, the dimer
 series contains all of them, but only to a certain (finite) order.
 Let us  mention however
that at present the dimer series expansion technique              
 has not yet been
 extended to calculate  two-particle properties, such as bound state
 energies and their contribution into the structure factor \cite{raj}.
  This fact makes it  worthwhile, in our view,  to  develop and further
 refine the technique, presented in this work, especially concerning
 the two-particle properties of quantum spin models with dimerization, of
 which the AHC is a particular example.      

\section{acknowledgements}
We are very grateful to Zheng Weihong, Rajiv Singh, and Anders Sandvik for 
numerous stimulating discussions. This work was supported by a grant from the Australian Research Council.

\vspace{0.5cm}

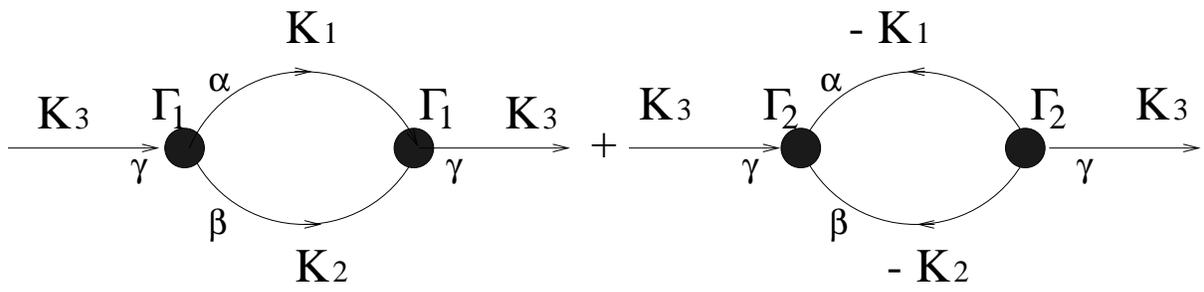
\begin{figure}
\caption{
Diagrams representing (second-order) corrections to the  one-particle 
normal self-energy due to the three-particle scattering $H_{3}$.} 
\label{fig1}
\end{figure}

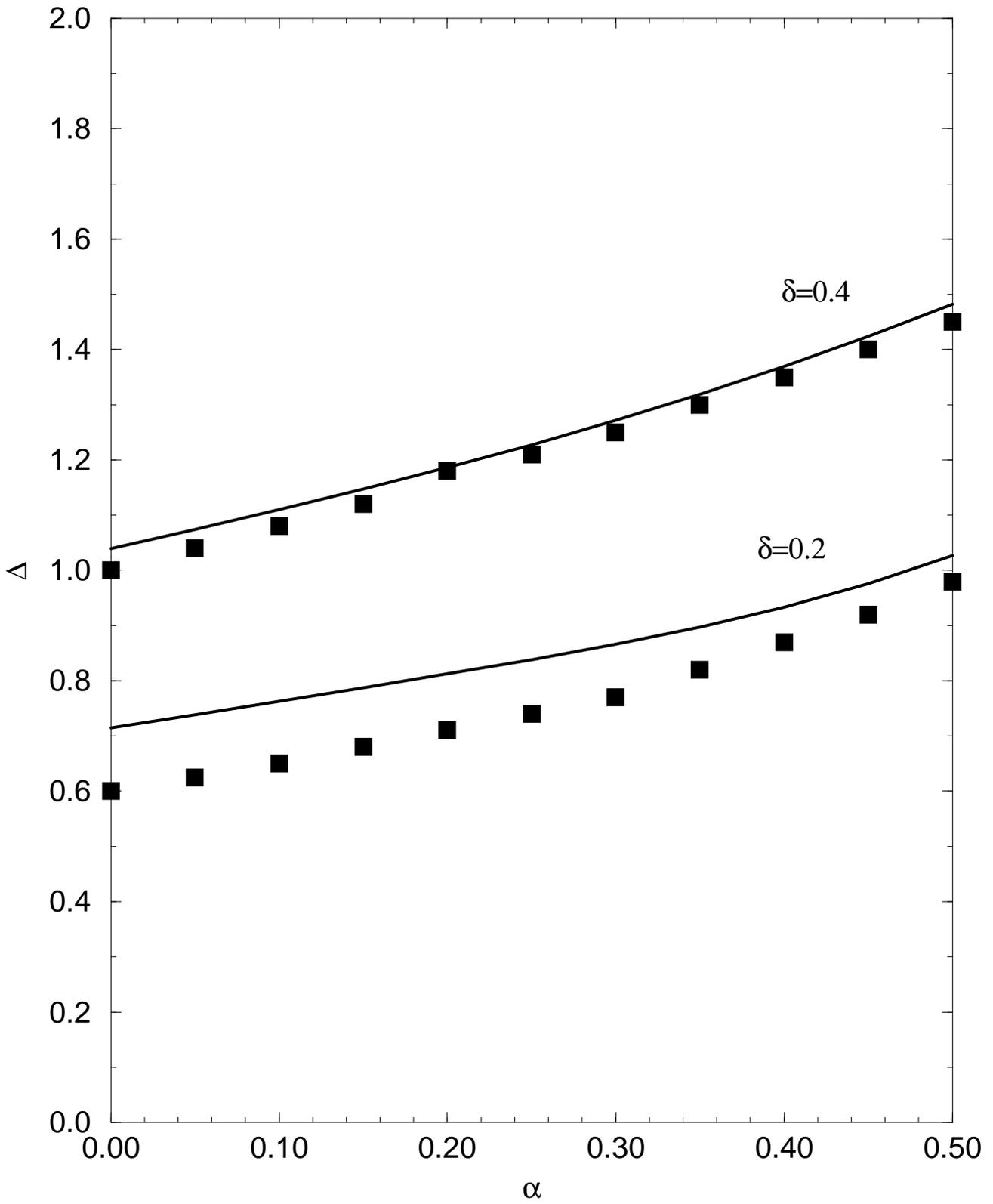
\begin{figure}
\caption{ 
The elementary triplet gap, 
$\Delta=\Omega_{k=0}/J$ versus $\alpha$ for 
$\delta=0.2, 0.4$. The squares  represent Exact Diagonalization  data \protect\cite{Bouzezar,Barnes}.}
\label{fig2}
\end{figure}

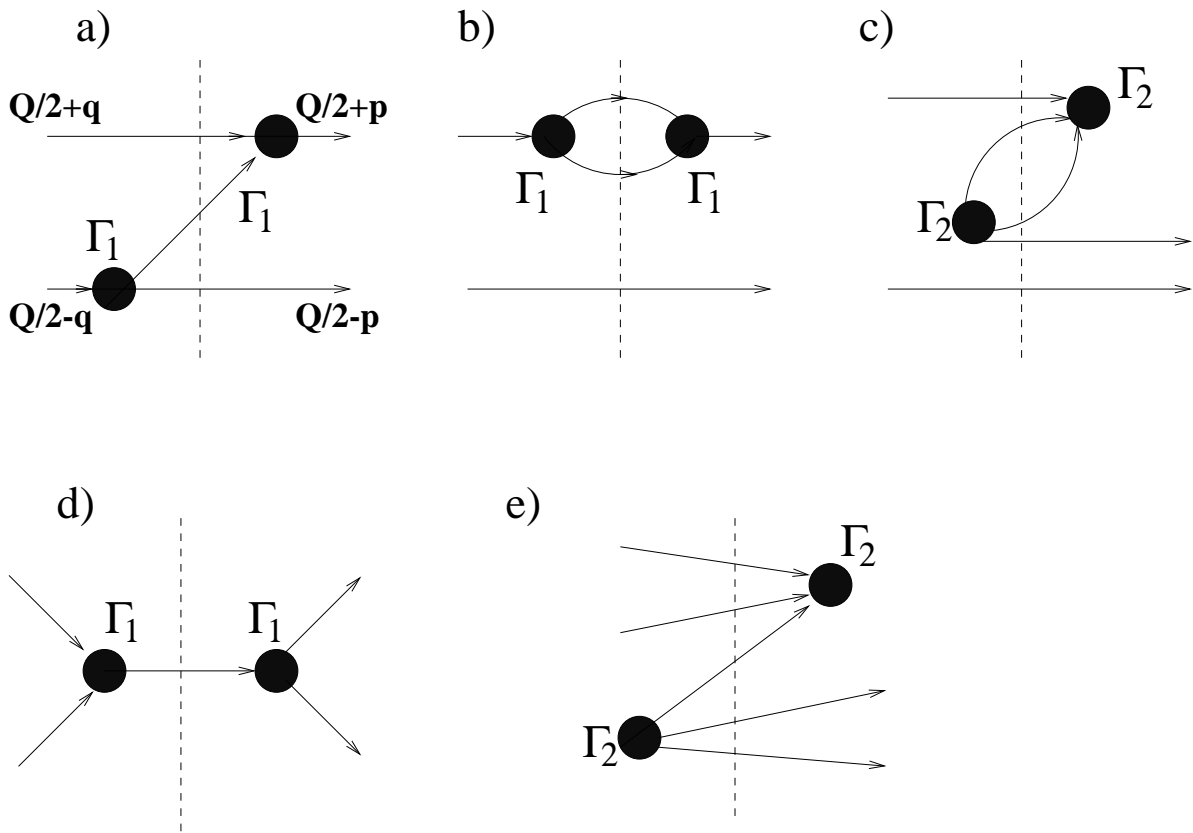
\begin{figure}
\caption{
Diagrams representing  the three-particle 
scattering contributions to the two-particle bound states. a),b),c) for the singlet bound state, and a),b),c),d),e) for the triplet bound state.}
\label{fig3}
\end{figure}
   
\begin{figure}
\caption{
 Energy spectrum, $\Omega_q/J$, for a) $\delta=0.2,\alpha=0.3$ 
(squares are Exact Diagonalization data \protect\cite{Bouzezar,Barnes}), 
b) $\delta=0.6,\alpha=0.0$, c) $\delta=0.6,\alpha=0.15$,
 d) $\delta=0.6,\alpha=0.2$, e) $\delta=0.8, \alpha=0.0$, f) $\delta=0.8, \alpha=0.3$. The solid lines are the elementary triplet 
spectrum (lower curve) and the lower edge of the two-triplet 
continuum (upper curve). The dashed and dot-dashed lines represent 
respectively the triplet, $S=1$, and singlet, $S=0$, bound states.}
\label{fig4}
\end{figure}

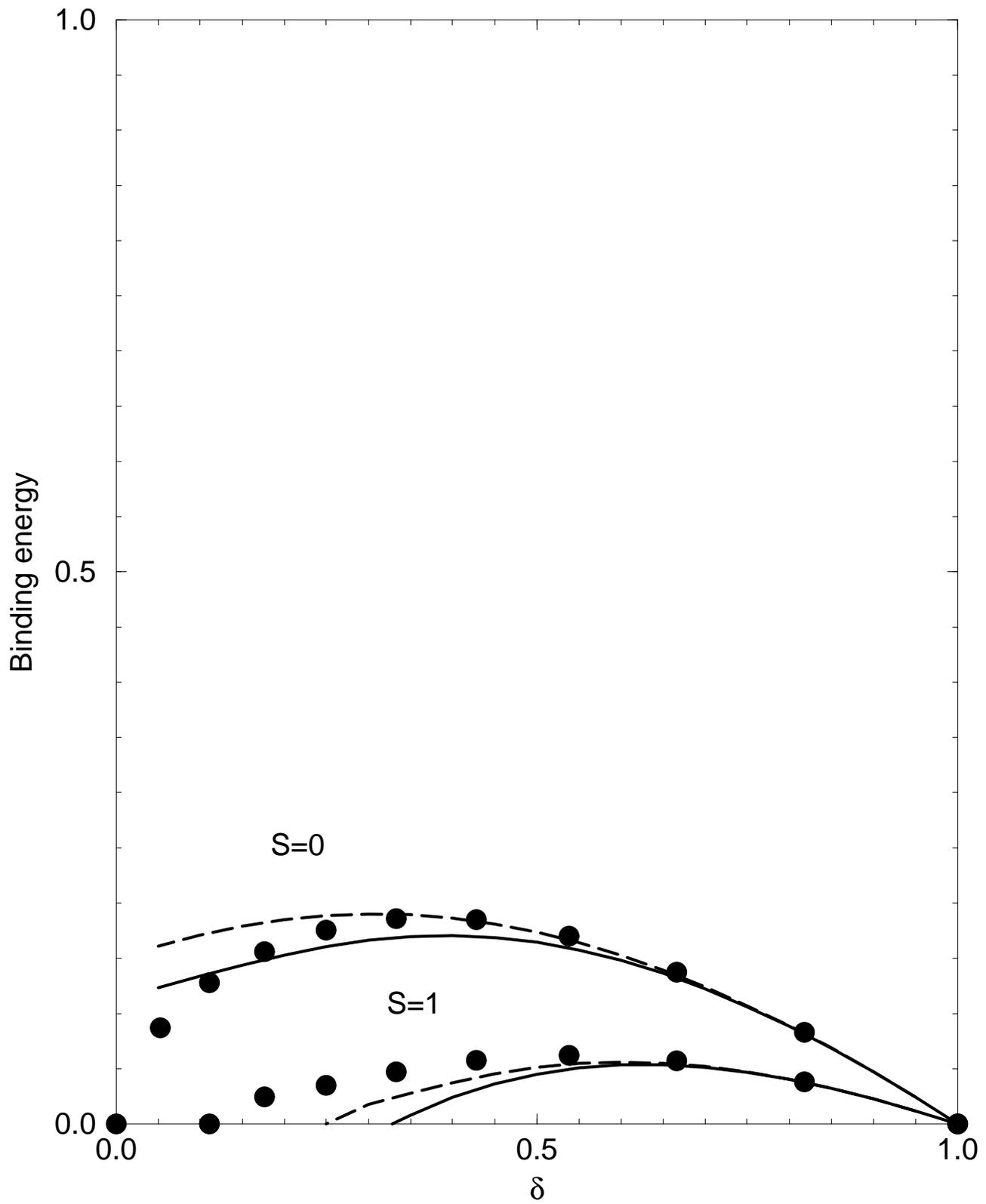
\begin{figure}
\caption{
 Binding energies, $\varepsilon^{S,T}/J$, of the two-particle singlet, $S=0$, and triplet, $S=1$,
 bound states  
 at $Q=\pi/2$, $\alpha=0$ versus $\delta$,  compared with the  
numerical Exact Diagonalization results (filled circles) \protect\cite{Barnes}. The solid lines represent the  calculations  
where 
the  three particle scattering $H_3$ is treated in  
simple perturbation theory, and the dashed lines represent the   self-consistent 
inclusion of  $H_3$ into the integral equation \protect(\ref{bs}). }
\label{fig5}
\end{figure}

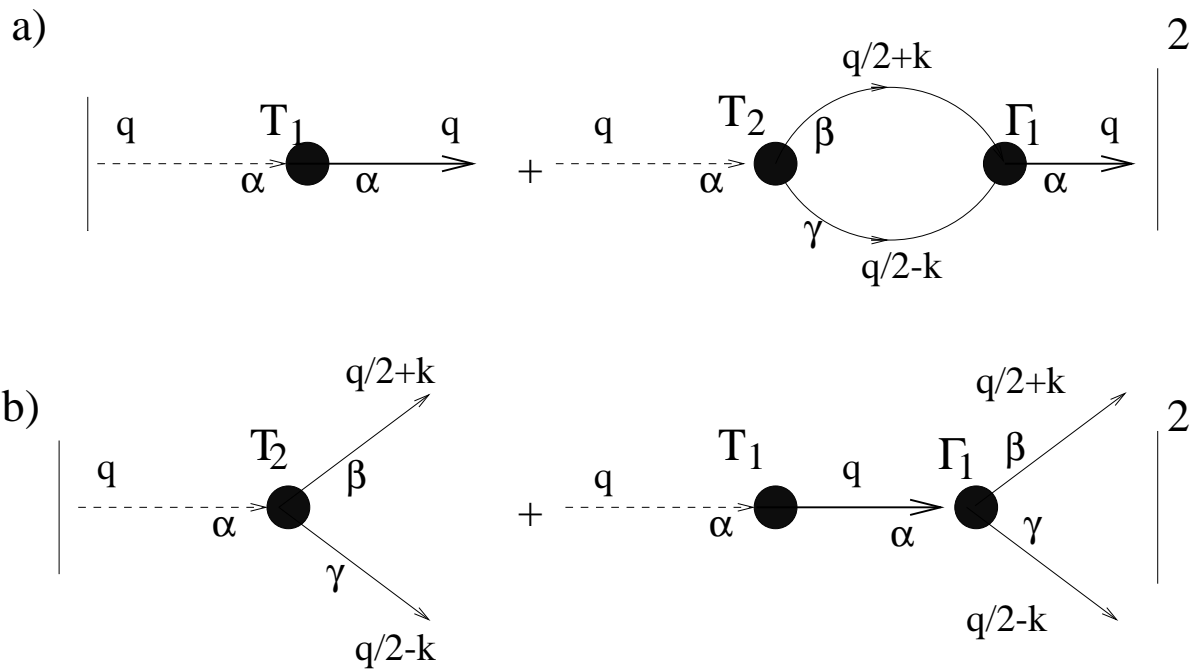
\begin{figure}
\caption{
 Diagrams for the  structure factors of a) 
the elementary triplet, and  b)  the two-particle triplet. }
\label{fig6}
\end{figure}

\begin{figure}
\caption{
 Structure factors $S_1(q)$ and $S_2(q)$ of the elementary  and
 the two-particle triplet:  
a) along the disordered line $1=\delta+2\alpha$ ($\lambda=0$),\\ 
b) for $\delta=0.4,\alpha=0.2,0.4$. The solid lines are the results of the  
calculations using Eqs.(\ref{str1},\ref{str2}).  
The symbols connected by dotted lines are the dimer series expansions results \protect\cite{Singh} for $S_1(q)$. }
\label{fig7}
\end{figure} 


\begin{thebibliography}{99}
\bibitem{remark} Strictly speaking there are other diagrams,
such as the "rainbow" contribution to the anomalous self-energy,
 which are of the same order in the density. These additional 
 contributions however are quite small, as discussed in
 Ref.\cite{Kotov}, and we neglect them here.
\end{thebibliography}
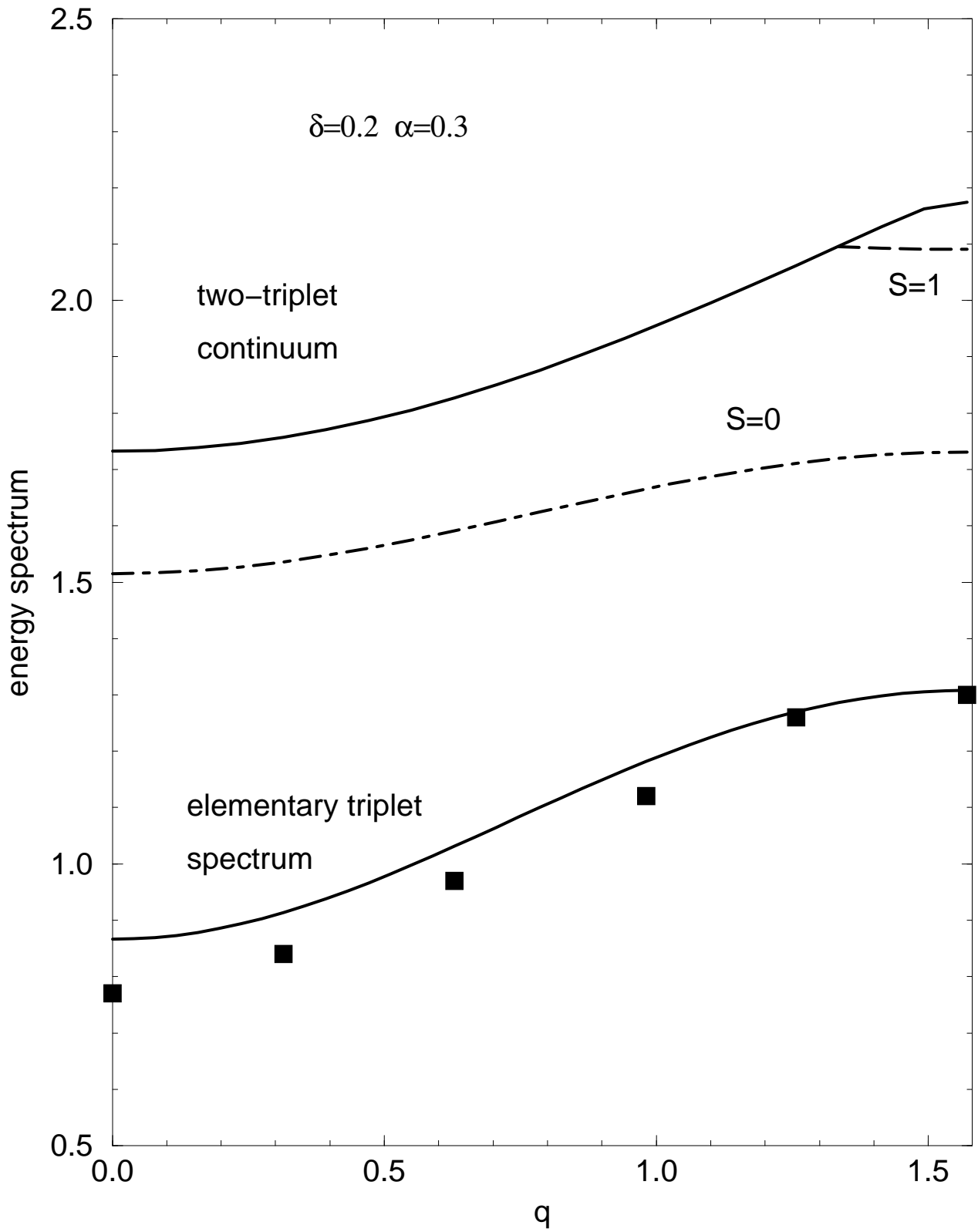
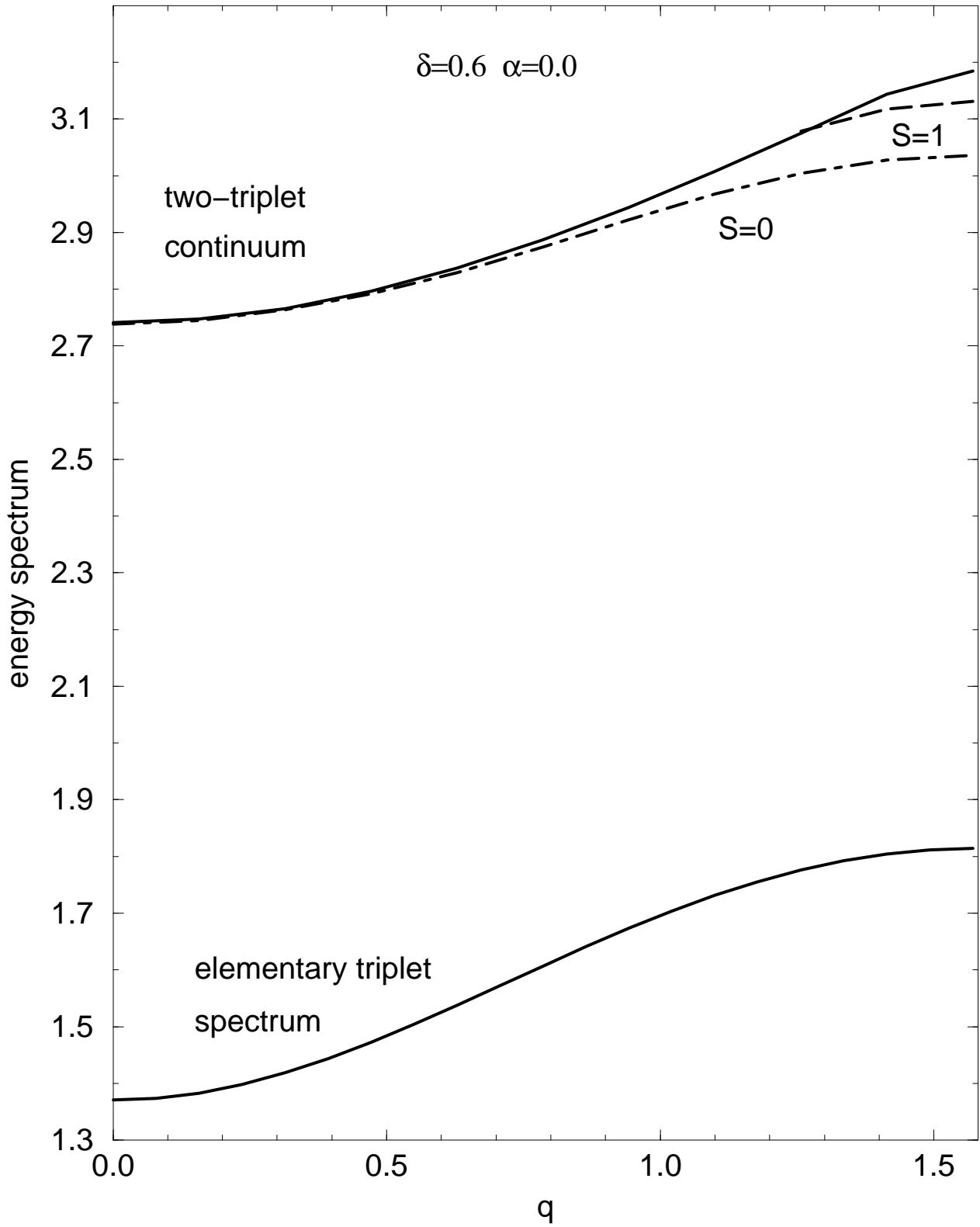
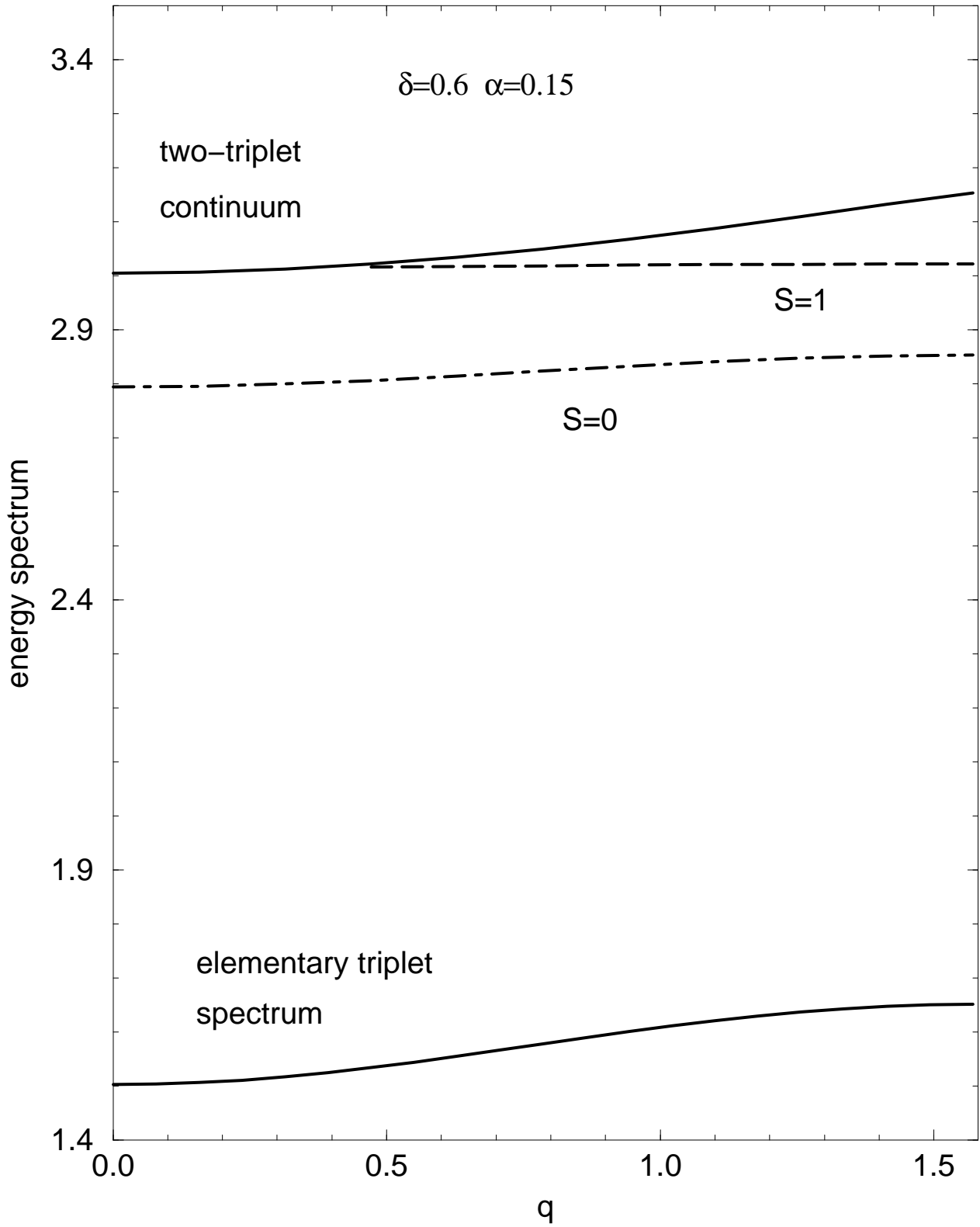
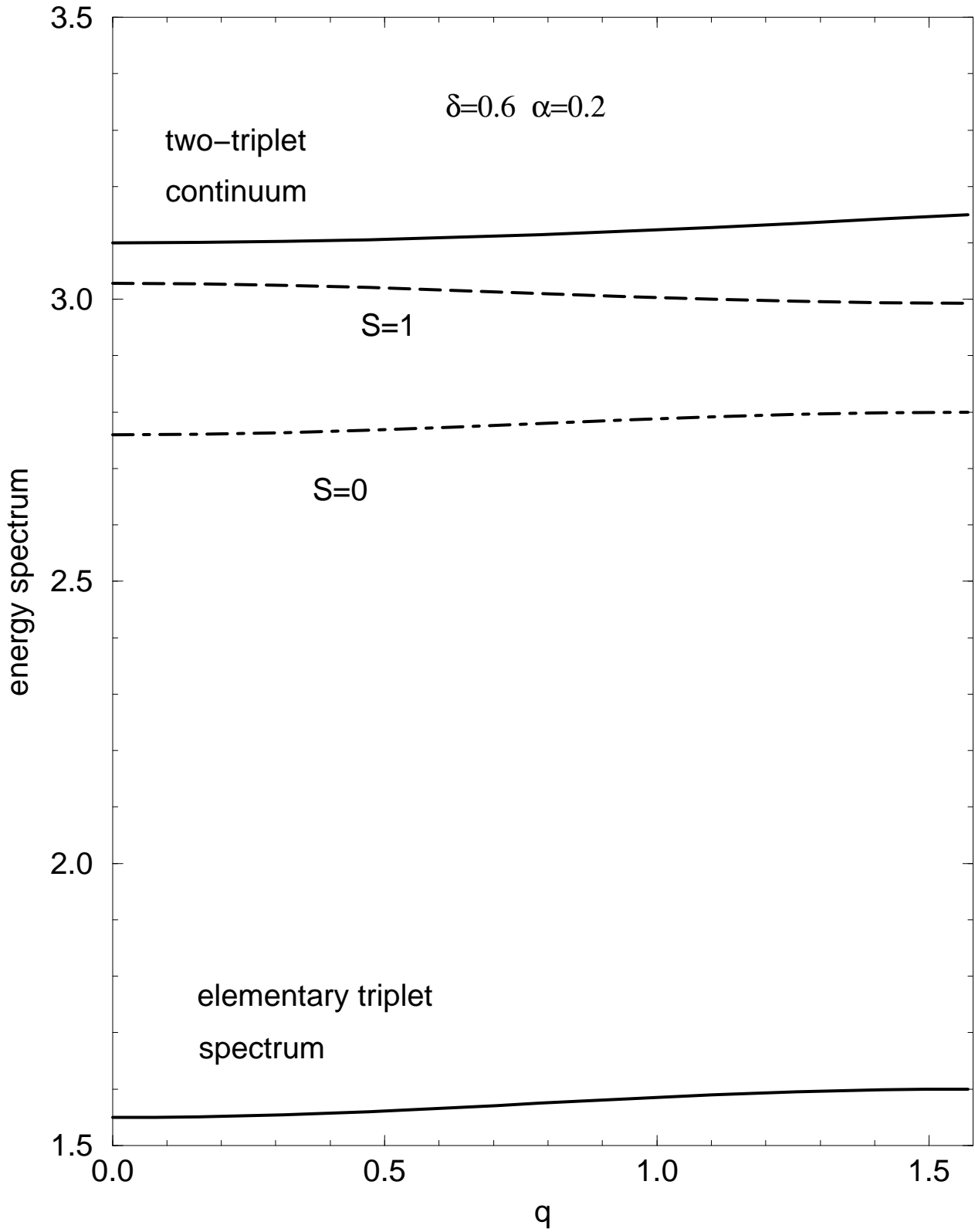
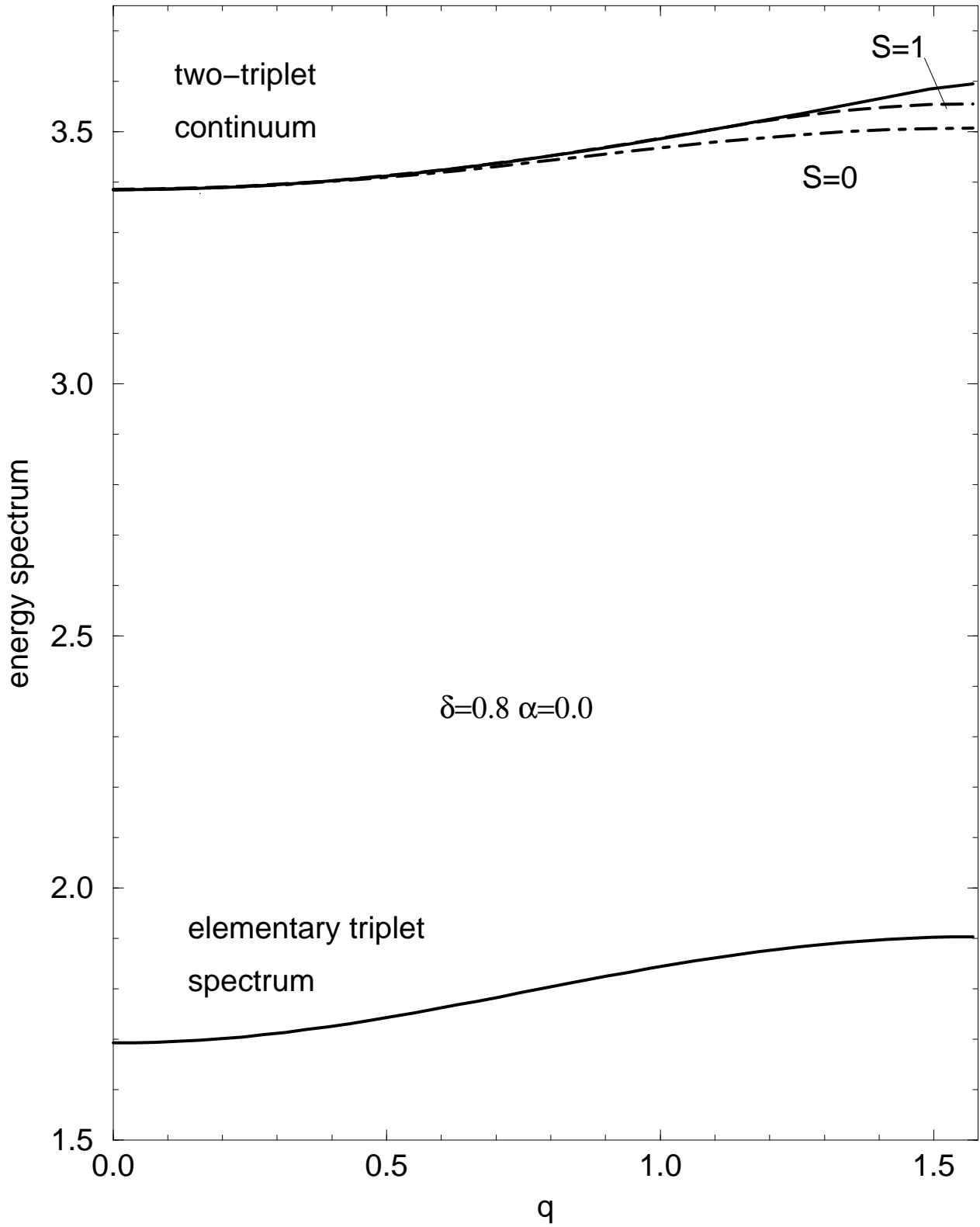
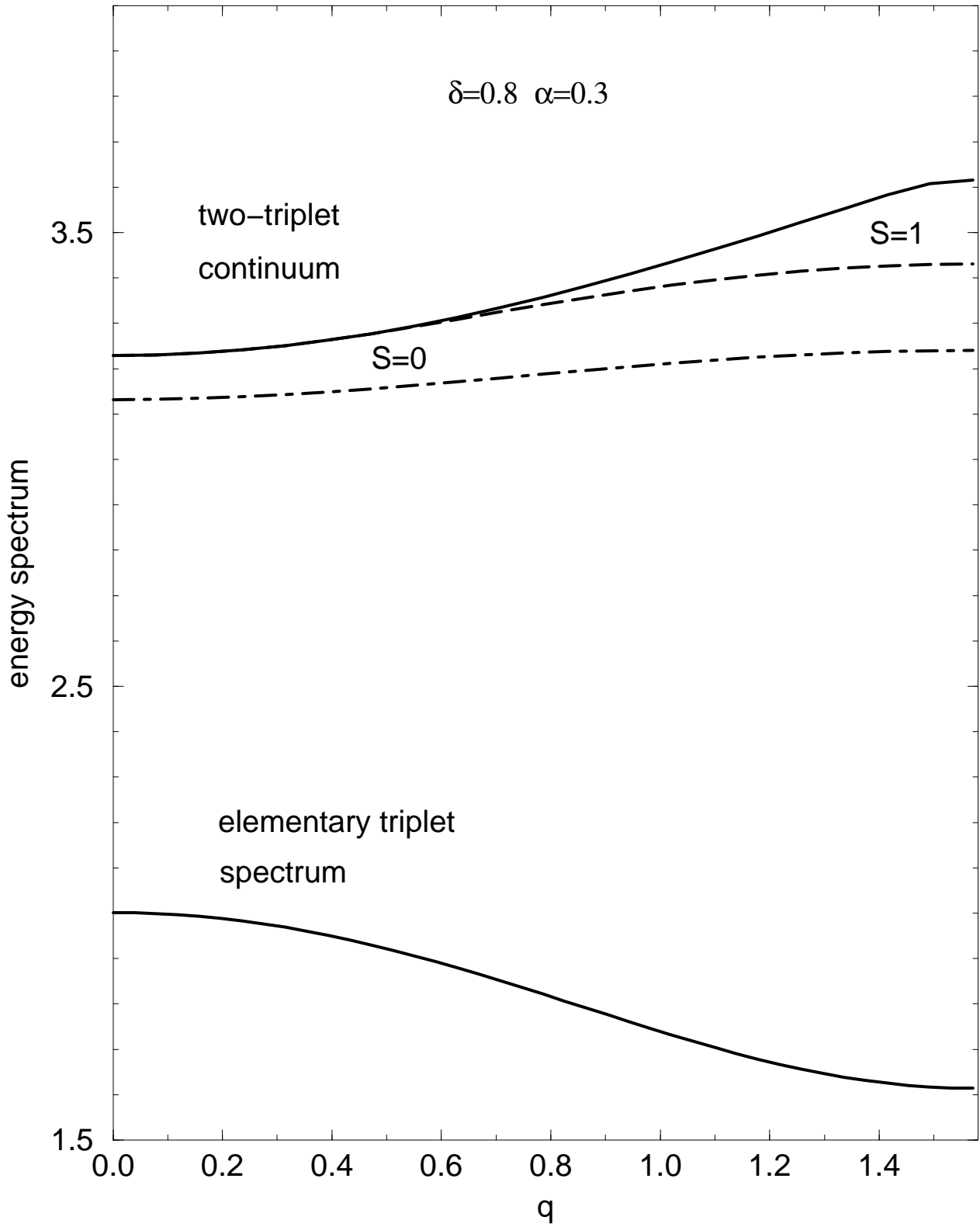
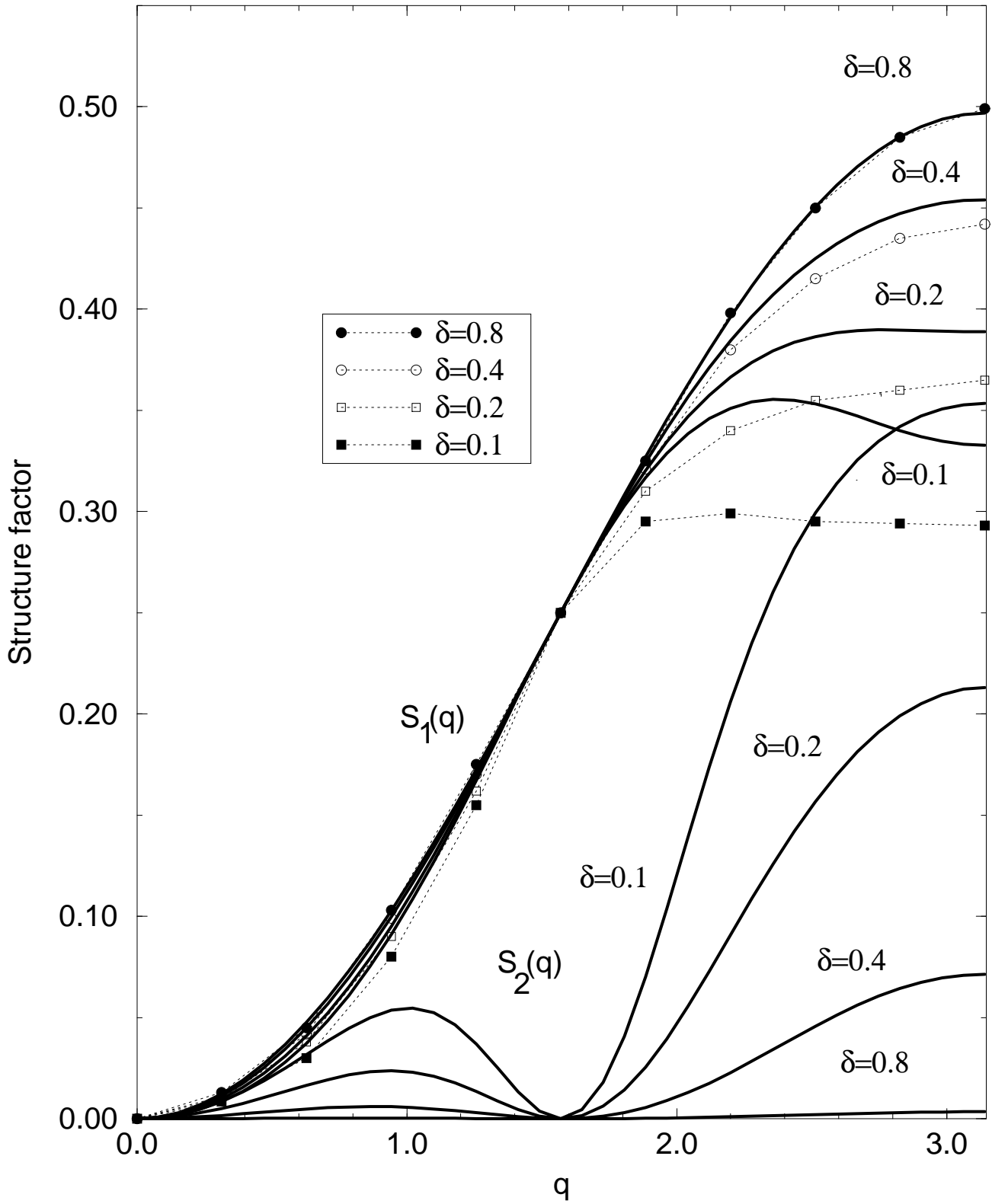
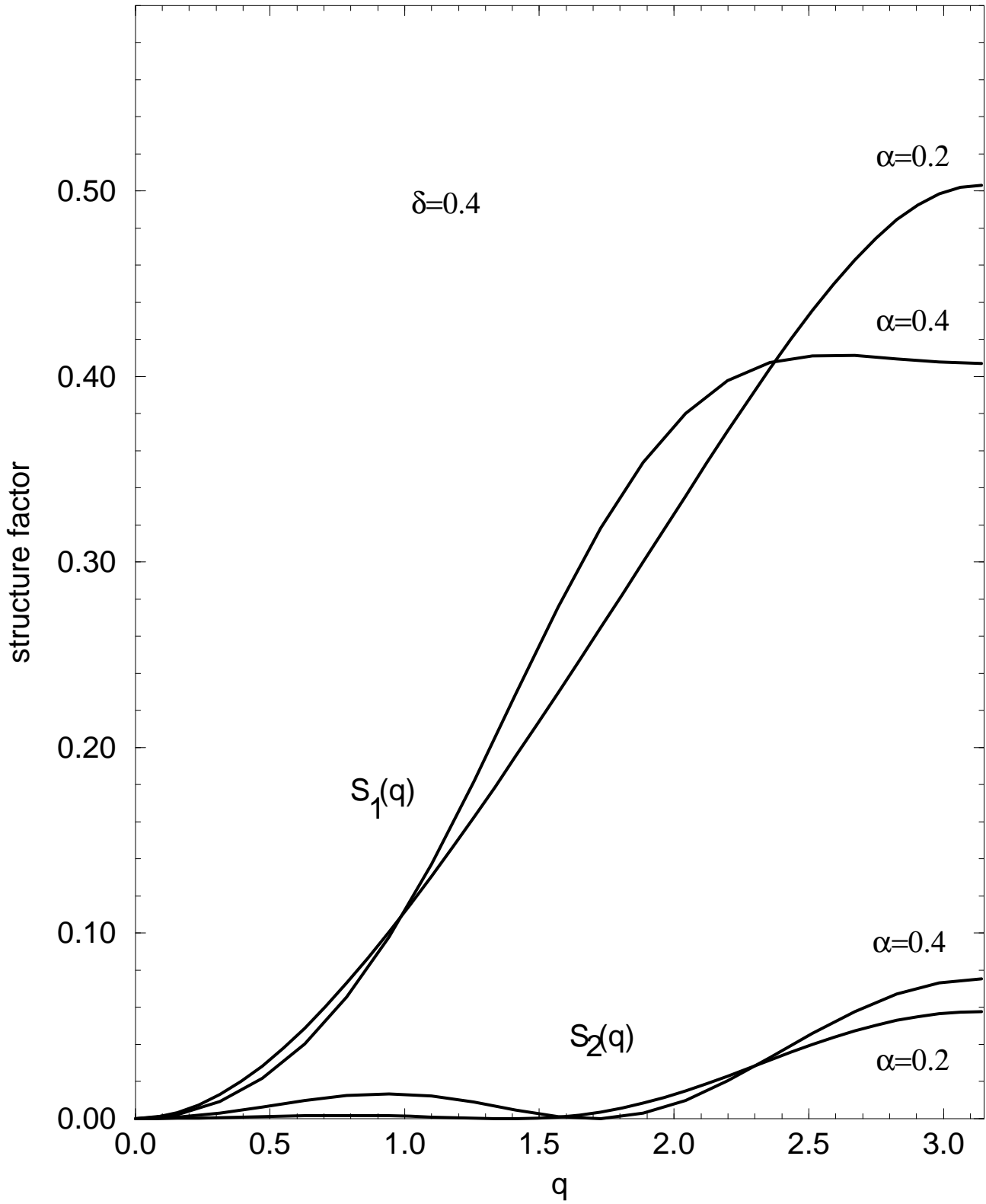
\end{document}